\documentstyle[amssymb,prb,aps]{revtex}

\begin{document}
\draft
\title{Vortex charge in mesoscopic superconductors}
\author{S.\ V.\ Yampolskii\cite{perm}, B. J.\ Baelus, F.\ M.\ Peeters\cite{aut2}}
\address{Departement Natuurkunde, Universiteit Antwerpen (UIA),\\
Universiteitsplein 1, B-2610 Antwerpen, Belgium}
\author{J.\ Kol\'{a}\v{c}ek}
\address{Institute of Physics, ASCR, Cukrovarnick\'{a} 10, Prague 6, Czech Republic}
\date{\today }
\maketitle

\begin{abstract}
The electric charge density in mesoscopic superconductors with circular
symmetry, i.e. disks and cylinders, is studied within the phenomenological
Ginzburg-Landau approach. We found that even in the Meissner state there is
a charge redistribution in the sample which makes the sample edge become
negatively charged. In the vortex state there is a competition between this
Meissner charge and the vortex charge which may change the polarity of the
charge at the sample edge with increasing magnetic field. It is shown
analytically that in spite of the charge redistribution the mesoscopic
sample as a whole remains electrically neutral.
\end{abstract}


\section{INTRODUCTION}

Recently it was predicted that the core of an Abrikosov vortex in bulk
type-II superconductors is charged~\cite{charge1,charge2,charge3}. This
effect occurs because of the difference of the chemical potential in the
superconducting versus normal state. Such a change in chemical potential
results in a redistribution of the electrons in the region near the vortex
core and culminates in a charging of the vortex core when the superconductor
is in the mixed state~\cite{charge2,charge3}.

In the present paper we investigate this phenomenon in mesoscopic
superconductors. A mesoscopic sample has a typical size which is comparable
to the coherence length $\left( \xi \right) $ and the magnetic field
penetration length $\left( \lambda \right) $. The behaviour of such
structures in an external magnetic field $(H)$\ is strongly influenced by
the sample shape~\cite{Mosch1,Schw1}\ and may lead to various
superconducting states and different phase transitions between them. Jumps
in magnetization were observed when the applied magnetic field or
temperature $(T)$ are varied~\cite{Geim1}.

A number of earlier works studied the geometry dependent magnetic response
of mesoscopic superconductors: i)~disk shape samples~\cite
{Deo1,Schw2,Schw3,Schw4,Pal,Akkerm,Deo2}; ii)~infinitely long cylinders~\cite
{cyl1,Zharkov}; iii)~ring-like structures~\cite{ring1,Ben1}\ and more
complicated geometries~\cite{FK}. Theoretical studies of mesoscopic
superconductors are based on the phenomenological Ginzburg-Landau (GL)
theory~\cite{book1}\ which successfully describes mesoscopic samples in a
wide $H$-$T$ region. In particular, it has been shown that in mesoscopic
samples (disks or cylinders) surrounded by a vacuum or an insulator two
kinds of superconducting states can exist. Firstly, there is a circular
symmetric state with a fixed value of angular momentum, called giant vortex.
The observed magnetization jumps correspond to first order phase transitions
between giant vortices with different angular momentum~\cite{Deo1,Schw2}.
Secondly, in samples with a sufficiently large radius multi-vortex
structures~\cite{Schw3} can nucleate\thinspace\ which are the analogue of
the Abrikosov flux line lattice in a bulk superconductor. These states can
be represented as a mixture of giant vortex states with different angular
momentum. For multi-vortex states it is also possible to introduce an
effective total angular momentum, which is nothing else then the number of
vortices in the disk, i.e. the vorticity. With changing the magnetic field
there is a second order phase transition between the multi-vortex and the
giant vortex state~\cite{Schw3,Schw4,Schw5}.

It is expected, that the charge distribution in such mesoscopic samples may
be appreciably altered due to the presence of a boundary. Furthermore,
screening currents near the boundary of the sample will also lead to a
redistribution of charge and consequently even in the Meissner state there
will be a non-uniform distribution in the sample. In a certain sense, this
case can be viewed as a vortex turned inside out, i.e. with its core at
infinity. In the presence of vortices there will be an interplay between the
Meissner charge and the previously studied~\cite{charge1,charge2,charge3}\
vortex charge.

The present paper is organized as follows. In Sec.~II we give the necessary
theoretical formalism on which our numerical results are based. In Sec.~III
we investigate the charge distribution in both the Meissner state and the
giant vortex states. Then the charge distribution in the multi-vortex state
is discussed for thin disks (Sec.~IV). Our results are summarized in Sec.~V.
In the Appendix we present the proof of electrical neutrality in a
mesoscopic superconductor of general shape.

\section{THEORETICAL\ APPROACH}

We consider a mesoscopic superconducting sample of circular symmetry with
radius $R$ and thickness $d$ surrounded by an insulating medium. The
external magnetic field $\overrightarrow{H}=(0,0,H)$ is uniform and directed
normal to the superconductor plane. The starting point of our analysis is
that the rotating motion of Cooper pairs around the vortex core leads to a
spatial redistribution of charge carriers, which generate the electrostatic
potential~\cite{charge1,charge2,charge3} 
\begin{equation}
\varphi \left( \overrightarrow{r}\right) =\varphi _{0}\left( \left| \psi
\left( \overrightarrow{r}\right) \right| ^{2}-1\right) ,  \label{pot1}
\end{equation}
where $\psi \left( \overrightarrow{r}\right) $\ is the dimensionless
superconducting order parameter normalized so that $\left| \psi \left( 
\overrightarrow{r}\right) \right| ^{2}$ is measured in units of the Cooper
pair density in a bulk superconductor. The amplitude $\varphi _{0}$ is
different in different approaches. We use $\varphi _{0}=\left| {\alpha }%
\right| {/}2e$ as proposed in Ref.~\cite{charge3}. A three-times smaller
value $\varphi _{0}=\left| {\alpha }\right| {/}6e$ was used in Ref.~\cite
{charge1}. From the theory in Ref.~\cite{charge2} it approximately follows $%
\varphi _{0}=\left( \left| \alpha \right| /2\pi ^{2}e\right) \left(
dT_{c}/d\ln \epsilon _{F}\right) $, where $dT_{c}/d\ln \epsilon _{F}\approx
\ln \left( \hbar \omega _{D}/k_{B}T_{c}\right) \sim 1$ to 10. Thus all
approaches yield an electrostatic potential of a similar magnitude.

The distribution of the corresponding charge density $q\left( 
\overrightarrow{r}\right) $ is obtained from the Poisson equation~\cite{tamm}
\begin{equation}
4\pi q\left( \overrightarrow{r}\right) =-\overrightarrow{\nabla }^{2}\varphi
\left( \overrightarrow{r}\right) .  \label{Poiss}
\end{equation}

The Cooper pair density $\left| \psi \left( \overrightarrow{r}\right)
\right| ^{2}$ is determined from a solution of the system of two coupled
non-linear GL equations for the superconducting order parameter, $\psi (%
\overrightarrow{r})$, and the magnetic field (or vector potential $%
\overrightarrow{A}(\overrightarrow{r})$) 
\begin{equation}
\left( -i\overrightarrow{\nabla }-\overrightarrow{A}\right) ^{2}\psi =\psi
-\psi \left| \psi \right| ^{2},  \label{GL1dim}
\end{equation}
\begin{equation}
\kappa ^{2}\overrightarrow{\nabla }\times \overrightarrow{\nabla }\times 
\overrightarrow{A}=\overrightarrow{j},  \label{GL2dim}
\end{equation}
where the density of the superconducting current $\overrightarrow{j}$ is
given by 
\begin{equation}
\overrightarrow{j}=\frac{1}{2i}\left( \psi ^{\ast }\overrightarrow{\nabla }%
\psi -\psi \overrightarrow{\nabla }\psi ^{\ast }\right) -\left| \psi \right|
^{2}\overrightarrow{A}.  \label{GL3}
\end{equation}
Here $\overrightarrow{r}=(\overrightarrow{\rho },z)$ is the
three-dimensional position in space. Due to the circular symmetry of the
sample we use cylindrical coordinates: $\rho $ is the radial distance from
the disk center, $\theta $ is the azimuthal angle and the $z$-axis is taken
perpendicular to the disk plane, where the disk lies between $z=-d/2$ and $%
z=d/2$. For $d\rightarrow \infty $ we obtain the cylinder geometry. All
distances are measured in units of the coherence length $\xi =\hbar /\sqrt{%
2m^{\ast }\left| \alpha \right| }$ ($m^{\ast }=2m$ is the mass of the Cooper
pair), the vector potential in $c\hbar /2e\xi $, the magnetic field in $%
H_{c2}=c\hbar /2e\xi ^{2}=\kappa \sqrt{2}H_{c},$ where $H_{c}$ is\ the
thermodynamical critical field, the superconducting current in $%
j_{0}=cH_{c}/2\pi \xi $, and $\kappa =\lambda /\xi $ is the GL parameter.

Eqs.~(\ref{GL1dim}-\ref{GL3}) has to be supplemented by boundary conditions
for $\psi (\overrightarrow{r})$\ and $\overrightarrow{A}(\overrightarrow{r})$%
. For the superconducting condensate it can be written as~\cite{book1} 
\begin{equation}
\left. \overrightarrow{n}\cdot \left( -i\overrightarrow{\nabla }-%
\overrightarrow{A}\right) \psi \right| _{S}=0,  \label{BCdim}
\end{equation}
where $\overrightarrow{n}$ is the unit vector normal to the sample surface.
The boundary condition for the vector potential has to be taken far away
from the superconductor where the magnetic field becomes equal to the
external applied field $H$%
\begin{equation}
\left. \overrightarrow{A}\right| _{r\longrightarrow \infty }=\frac{1}{2}%
H\rho \overrightarrow{e}_{\theta },
\end{equation}
where $\overrightarrow{e}_{\theta }$ denotes the azimuthal direction.

The free energy of the superconducting state, measured in $%
F_{0}=H_{c}^{2}V/8\pi $ units, is determined by the expression 
\begin{equation}
F=\frac{2}{V}\left\{ \int dV\left[ -\left| \psi \right| ^{2}+\frac{1}{2}%
\left| \psi \right| ^{4}+\left| -i\overrightarrow{\nabla }\psi -%
\overrightarrow{A}\psi \right| ^{2}+\kappa ^{2}\left( \overrightarrow{h}%
\left( \overrightarrow{r}\right) -\overrightarrow{H}\right) ^{2}\right]
\right\}  \label{F1}
\end{equation}
with the magnetic field 
\[
\overrightarrow{h}\left( \overrightarrow{r}\right) =\overrightarrow{\nabla }%
\times \overrightarrow{A}(\overrightarrow{r}). 
\]

Here we will consider three different geometries for the superconductor:
I)~an infinitely long cylinder; II)~a thin disk with finite thickness $%
d\lesssim \lambda ,\xi $; and III)~an infinitely thin disk, i.e. $%
d\rightarrow 0$. In all three cases the superconducting order parameter does
not depend on $z$. In case I it obviously follows from the sample geometry.
In cases~II and~III it was found~\cite{Schw2} that the dependence of $\psi (%
\overrightarrow{r})$\ on $z$ is very slow. This allows us to average $\psi (%
\overrightarrow{r})$\ over the sample thickness and to solve the problem for
the two-dimensional problem for the order parameter $\psi \left( \rho
,\theta \right) $. However, the magnetic field in case II has a $z$%
-dependence, which is responsible for the demagnetization effect.\ For both
cases I and II we solve the problem numerically by the method proposed in
Ref.~\cite{Schw2}. For thin mesoscopic disks we use the results of Refs.~ 
\cite{Schw3,enhance} which allowed us to solve the problem semi-analytically.

\section{CHARGE\ IN\ THE\ MEISSNER\ AND\ THE\ GIANT VORTEX\ STATES}

Firstly, we consider the situation with a fixed value of the vorticity $L$.
The giant vortex state has cylindrical symmetry and consequently the order
parameter can be written as $\psi \left( \overrightarrow{\rho }\right)
=f\left( \rho \right) \exp \left( iL\theta \right) $. For a thin disk (case
III)) the order parameter is~\cite{enhance} 
\begin{equation}
\psi \left( \rho ,\theta \right) =\left( -\Lambda \frac{I_{2}}{I_{1}}\right)
^{1/2}f_{L}\left( \rho \right) \exp \left( iL\theta \right) ,  \label{psiGV}
\end{equation}
where 
\begin{equation}
f_{L}\left( \rho \right) =\left( \frac{H\rho ^{2}}{2}\right) ^{L/2}\exp
\left( -\frac{H\rho ^{2}}{4}\right) M\left( -\nu ,L+1,\frac{H\rho ^{2}}{2}%
\right) ,  \label{fGV}
\end{equation}
\begin{equation}
I_{1}=\int\limits_{0}^{R}\rho d\rho \ f_{L}^{4}\left( \rho \right) ,\quad
I_{2}=\int\limits_{0}^{R}\rho d\rho \ f_{L}^{2}\left( \rho \right) ,
\end{equation}
\begin{equation}
\Lambda =H\left( 1+2\nu \right) -1.  \label{eig1}
\end{equation}
Here $M\left( a,b,y\right) $ is the Kummer function~\cite{book2} and the
value of $\nu $ is determined by the non-linear equation, which results from
the boundary condition~(\ref{BCdim}) 
\begin{equation}
\left( L-\frac{\Phi }{2}\right) M\left( -\nu ,L+1,\frac{\Phi }{2}\right) -%
\frac{\nu \Phi }{L+1}M\left( -\nu +1,L+2,\frac{\Phi }{2}\right) =0.
\label{BCN}
\end{equation}
Here $\Phi =HR^{2}$ is the magnetic flux through the disk in the absence of
any flux expulsion. Using (\ref{psiGV}-\ref{eig1}) we can derive explicitly
the charge distribution 
\begin{eqnarray}
q\left( \rho \right) &=&4\Lambda \frac{I_{2}}{I_{1}}\left( \frac{H}{2}%
\right) ^{L}\rho ^{2\left( L-1\right) }\exp \left( -\frac{H\rho ^{2}}{2}%
\right)  \label{chGV} \\
&&\times \left\{ L^{2}M^{2}\left( -\nu ,L+1,\frac{H\rho ^{2}}{2}\right) -%
\frac{H\rho ^{2}}{2}\left( 2L+1\right) M\left( -\nu ,L+1,\frac{H\rho ^{2}}{2}%
\right) \right.  \nonumber \\
&&\times \left[ M\left( -\nu ,L+1,\frac{H\rho ^{2}}{2}\right) +\frac{2\nu }{%
L+1}M\left( -\nu +1,L+2,\frac{H\rho ^{2}}{2}\right) \right]  \nonumber \\
&&+\left( \frac{H\rho ^{2}}{2}\right) ^{2}\left[ M^{2}\left( -\nu ,L+1,\frac{%
H\rho ^{2}}{2}\right) +\frac{4\nu }{L+1}M\left( -\nu ,L+1,\frac{H\rho ^{2}}{2%
}\right) \right.  \nonumber \\
&&\times M\left( -\nu +1,L+2,\frac{H\rho ^{2}}{2}\right) +\frac{2\nu ^{2}}{%
\left( L+1\right) ^{2}}M^{2}\left( -\nu +1,L+2,\frac{H\rho ^{2}}{2}\right) 
\nonumber \\
&&\left. \left. -\frac{2\nu \left( \nu +1\right) }{\left( L+1\right) \left(
L+2\right) }M\left( -\nu ,L+1,\frac{H\rho ^{2}}{2}\right) M\left( -\nu
+2,L+3,\frac{H\rho ^{2}}{2}\right) \right] \right\} ,  \nonumber
\end{eqnarray}
where the charge density is measured in units of $q_{0}=\hbar ^{2}/\left(
16\pi m^{\ast }e\xi ^{4}\right) =\left( a_{B}/32\pi \xi ^{4}\right) e$ ($%
a_{B}=\hbar ^{2}/me^{2}$ is the Bohr radius). For Al we have $\xi =250$ nm~ 
\cite{Geim1} which results into $q_{0}/e\simeq 1.3\cdot 10^{-13}$ nm$^{-3}$.

\subsection{MEISSNER\ STATE}

First, we study the Meissner state, i.e. $L=0$. The radial dependences of
the Cooper pair density (or the corresponding distribution of the potential $%
\varphi \left( \rho \right) $ (see Eq.~(\ref{pot1})), the charge density $%
q(\rho )$\ and the density of superconducting current $j\left( \rho \right) $%
\ are shown in Figs.~\ref{L0}(a-c) for the cases of cylinder, finite disk
and very thin disk, respectively, for different values of the applied field.
Fig.~\ref{L0}(d) shows these dependencies for a cylinder at the field $%
H=0.52H_{c2}$ but for different $\kappa $ values. Due to the finite radial
size of the samples all distributions are inhomogeneous along the radius of
the sample. The Cooper pair density is maximum at the center and decays
towards the sample edge. As a result, in the center of the sample there is a
region of positive charge while near the edge a negative ``screening''
charge $q_{scr}$\ is created. To avoid confusion let us note that for
simplicity we write ``positive charge'' instead of ``charge of the same sign
as is the sign of the dominant charge carriers''. In the cylinder the Cooper
pair density decreases with increasing field, while both the screening
superconducting current and the charge polarization monotoneously increase.
The behavior for the disks is more complicated. For small fields the picture
is very similar to the one of the cylinder. But for fields where the
Meissner state becomes metastable (i.e., $H/H_{c2}>0.32$ for the parameters
used in Fig.~\ref{L0}(b)) the screening charge becomes maximal. The positive
charge is pulled further towards the center of the disk and the screening
charge region expands. Notice that now the maximum of the positive charge
decreases with field and also at the surface its absolute value decreases.
For a cylinder, at fixed field but with increasing $\kappa $ (Fig.~\ref{L0}%
(d))\ its charge distribution behaves similar as for fixed $\kappa $\ and
increasing magnetic field (see Fig.~\ref{L0}(a)).

Notice that even for $L=0$ when no vortex is present inside the
superconductor, there is still a non-uniform charge distribution. This
charge redistribution can be characterised by two quantities: i) the
distance $\rho ^{\ast }$ which separates the positive and negative charge
regions, and ii) the total screening charge $Q_{-,scr}$\ which is defined by
the integral over the region $V_{-}$ occupied by the negative charge: 
\begin{equation}
Q_{-,\;scr}=2\pi d\int\limits_{\left( \rho _{-}\right) }\rho q_{scr}\left(
\rho \right) d\rho .  \label{Qneg}
\end{equation}

The dependences of $\rho ^{\ast }\left( H\right) $ are shown in Fig.~\ref
{rstar}(a-c) for the cylinder, the finite disk and the thin disk,
respectively. In Fig.~\ref{rstar}(c) the open squares refer to the magnetic
field region where the $L=0$ state is metastable (the crossed circles will
be explained below). Notice that $\rho ^{\ast }$\ decreases with increasing
field and this decrease is more pronounced for thinner disks. The magnetic
field dependences of the absolute value $\left| Q_{-}\right| $\ of the
screening charge is shown in Fig.~\ref{quabs}(a-c) for the same geometries.
The screening charge $Q_{scr}$ increases with magnetic field, but for the
disk cases a local maximum is reached. This local maximum is reached for
fields where $\rho ^{\ast }$ starts to decrease more strongly. Taking $\xi
=250$ nm and $a_{B}=0.05$ nm we obtain that for the characteristic value of
screening charge 
\mbox{$\vert$}%
$Q_{-,\;scr}|\simeq 3\cdot 2\pi \xi ^{2}q_{0}d\simeq 8\cdot 10^{-6}e$ at the
field $H=0.3H_{c2}$\ for the Al disk with $d/\xi =0.2$ (the case II). For
strongly type-II superconductors which have a very small coherence length $%
\xi $\ this induced charge can be made orders of magnitude larger.

By direct integration of $q\left( \rho \right) $ over the sample surface one
can convince ourselves that the total charge per unit of sample length $%
Q=\oint d\theta \int\nolimits_{0}^{R}\rho q\left( \rho \right) d\rho =0$,
i.e. there is charge neutrality over the whole sample. But this fact can be
easy generalized analytically for states with any vorticity. The proof is
given in the Appendix.

\subsection{GIANT VORTEX STATE}

The same dependences as in Figs.~\ref{L0}(a-d) are shown in Figs.~\ref{L1}%
(a-d) for samples in the $L=1$\ vortex state. In this state $\left| \psi
\left( \rho \right) \right| ^{2}=0$ in the center of the vortex core located
in the centre of the sample. Notice, that for a cylinder the charge
distribution almost does not change with magnetic field. The reason is that
the external field affects the Cooper pair density only near the sample edge
region. This is different for the disk geometry, where large demagnetization
effects strongly influences the\ penetration of the magnetic field in the
disk which changes the vortex structure. The vortex core is negatively
charged and at small fields the positive charge outside the vortex extends
up to the border of the sample. The dependences of the position $\rho ^{\ast
}\left( H\right) $ and the size of the charge pile-up $\left| Q_{-}\right|
\left( H\right) $ for the $L=1$ state are shown in Figs.~\ref{rstar}(a-c)
and \ref{quabs}(a-c).

With increasing external magnetic field the screennig current at the sample
surface increases and makes the region near the border of the sample
negatively charged. In this case there exist two $\rho ^{\ast }$, where $%
q\left( \rho ^{\ast }\right) =0$. The core of the vortex is negatively
charged with total charge $\left| Q_{-,\;v}\right| $. Around this core there
is a ring of positive charge which compensates the charge of the vortex core
and the surface charge. Near the surface a ring of negative screening charge
exists with total charge $\left| Q_{-,\;scr}\right| $. The size of the
latter increases with increasing magnetic field. For disks, $\left|
Q_{-,\;scr}\right| $ reaches a local maximum after which it decreases in the
large magnetic field region; this is the region where the $L=1$ state is
unstable (see circles with crosses in Fig.~\ref{quabs}(c)).

Next we investigated the charge distribution for the vortex states with $L>1$
and we limited ourselves to the thin disk case. From Eq.~(\ref{chGV}) one
finds immediately that $q=0$ in the center of the disk when $L\geq 2$.
Consequently, the charge distribution in the vortex core has a ring shape.
To illustrate this we show in Figs.~\ref{L23}(a,b) the same dependences as
in Fig.~\ref{L1}(c) but now for $L=2$ and $3$, respectively. With the
exception of the core region the charge distribution for the giant vortex
states is qualitatively similar to the case $L=1$, and the charge on the
sample surface changes sign with increasing external magnetic field. Notice
also that the number of areas where the charge changes its sign does not
increase with $L$: it equals $2$ for small fields and increases to $3$ for
higher magnetic fields.

\section{VORTEX CHARGE\ IN\ THE\ MULTI-VORTEX\ STATE}

For sufficiently large radial size of the superconductor the giant vortex
state can break up into multi-vortices~\cite{Schw3,Schw4,Pal}. To explain
the physics we limit ourselves to the case of a thin disk. It was shown~\cite
{Schw4,Pal} that the order parameter of the multi-vortex state in general
can be viewed as a superposition of giant vortex states with different $%
L_{j} $%
\begin{equation}
\psi \left( \overrightarrow{\rho }\right)
=\sum\limits_{L_{j}=0}^{L}C_{L_{j}}f_{L_{j}}\left( \rho \right) \exp \left(
iL_{j}\theta \right) ,  \label{SP}
\end{equation}
where $L$ is now the value of the effective total angular momentum which
equals the number of vortices in the disk. For disks with not so large
radius $(3\div 5\;\xi )$ the order parameter of the multi-vortex state is
the superposition of only two states and is described by the expression~\cite
{enhance} 
\begin{equation}
\psi \left( \overrightarrow{\rho }\right) =C_{L_{1}}f_{L_{1}}\left( \rho
\right) \exp \left( iL_{1}\theta \right) +C_{L_{2}}f_{L_{2}}\left( \rho
\right) \exp \left( iL_{2}\theta \right) ,  \label{psimv}
\end{equation}
where 
\begin{eqnarray}
C_{L_{1}} &=&\left( \frac{-\Lambda _{L_{1}}A_{L_{2}}B_{L_{1}}+2\Lambda
_{L_{2}}A_{L_{1},L_{2}}B_{L_{2}}}{A_{L_{1}}A_{L_{2}}-4A_{L_{1},L_{2}}^{2}}%
\right) ^{1/2},  \label{Cmv1} \\
C_{L_{2}} &=&\left( \frac{-\Lambda _{L_{2}}A_{L_{1}}B_{L_{2}}+2\Lambda
_{L_{1}}A_{L_{1},L_{2}}B_{L_{1}}}{A_{L_{1}}A_{L_{2}}-4A_{L_{1},L_{2}}^{2}}%
\right) ^{1/2},  \nonumber
\end{eqnarray}

\begin{equation}
A_{L_{i}}=\frac{2\pi d}{V}\int\limits_{0}^{R}\rho d\rho \
f_{L_{i}}^{4}\left( \rho \right) ,\;A_{L_{1},L_{2}}=\frac{2\pi d}{V}%
\int\limits_{0}^{R}\rho d\rho \ f_{L_{1}}^{2}\left( \rho \right)
f_{L_{2}}^{2}\left( \rho \right) ,\;B_{L_{i}}=\frac{2\pi d}{V}%
\int\limits_{0}^{R}\rho d\rho \ f_{L_{i}}^{2}\left( \rho \right) ,
\label{Cmv2}
\end{equation}
and $f_{L_{i}}\left( \rho \right) $ and $\Lambda _{L_{i}}$ are determined by
Eqs.~(\ref{fGV}) and (\ref{eig1}), respectively. The charge density
distribution is then given by the expression 
\begin{eqnarray}
q\left( \rho ,\theta \right) &=&2C_{L_{1}}^{2}\left[ f_{L_{1}}^{\prime
2}\left( \rho \right) +f_{L_{1}}\left( \rho \right) f_{L_{1}}^{\prime \prime
}\left( \rho \right) +\frac{1}{\rho }f_{L_{1}}\left( \rho \right)
f_{L_{1}}^{\prime }\left( \rho \right) \right]  \label{chMV} \\
&&+2C_{L_{2}}^{2}\left[ f_{L_{2}}^{\prime 2}\left( \rho \right)
+f_{L_{2}}\left( \rho \right) f_{L_{2}}^{\prime \prime }\left( \rho \right) +%
\frac{1}{\rho }f_{L_{2}}\left( \rho \right) f_{L_{2}}^{\prime }\left( \rho
\right) \right]  \nonumber \\
&&+2C_{L_{1}}C_{L_{2}}\cos \left[ \left( L_{1}-L_{2}\right) \theta \right] %
\left[ f_{L_{1}}^{\prime \prime }\left( \rho \right) f_{L_{2}}\left( \rho
\right) +2f_{L_{1}}^{\prime }\left( \rho \right) f_{L_{2}}^{\prime }\left(
\rho \right) +f_{L_{1}}\left( \rho \right) f_{L_{2}}^{\prime \prime }\left(
\rho \right) \right.  \nonumber \\
&&\left. +\frac{1}{\rho }\left( f_{L_{1}}^{\prime }\left( \rho \right)
f_{L_{2}}\left( \rho \right) +f_{L_{1}}\left( \rho \right) f_{L_{2}}^{\prime
}\left( \rho \right) \right) -\frac{\left( L_{1}-L_{2}\right) ^{2}}{\rho ^{2}%
}f_{L_{1}}\left( \rho \right) f_{L_{2}}\left( \rho \right) \right] , 
\nonumber
\end{eqnarray}
where the prime denotes the derivative with respect to $\rho $. The explicit
expression is rather lengthy and is therefore not given here.

Earlier analyses have shown~\cite{Schw4,enhance,Ben2}\ that there exist two
kinds of multi-vortex states: i) stable configurations which correspond to a
minimum of the free energy; and ii) states which correspond to saddle points
of the free energy. The latter ones correspond to the energy barrier states
between states with different vorticity $L$\ and describe the penetration of
flux into the disk. Due to the transitions between the different $L$\ states
with increasing (or decreasing) external field some giant vortex states are
never realised (for example, such states for $L=0$ and $1$ correspond to the
crossed circles in Figs.~\ref{rstar}(c) and~\ref{quabs}(c)).

As an example, we consider a thin disk with $R/\xi =4.0$. The charge density 
$q\left( x,y\right) $ distribution over the disk for the different kinds of
multi-vortex states are shown in Figs.~\ref{mv_saddle} and~\ref{mv_stable}.
In Fig.~\ref{mv_saddle} this distribution is given for the saddle point
state ($1$:$2$) at the field $H=0.32H_{c2}$ and the contour plot of the
distribution of the Cooper pair density $\left| \psi \right| ^{2}$\ for this
state is shown at the bottom of the figure. The dark regions on the $\left|
\psi \right| ^{2}$ contour plot correspond to low Cooper pair density. The
same distributions for the stable multi-vortex state ($0$:$4$) at the field $%
H=0.75H_{c2}$\ are shown in Fig.~\ref{mv_stable}. One can see regions of
negative charge located at the vortex core and positive charge near the edge
of the sample. Using Eqs.~(\ref{psimv}-\ref{Cmv2}) it is easy to prove that
also in the multi-vortex state the disk is electrically neutral as a whole
(see also the Appendix).

\section{CONCLUSIONS}

We studied theoretically the redistribution of electrical charge in circular
mesoscopic superconducting samples with different shape, i.e. disks and
cylinders. The theory applies for intermediate temperatures where the
Ginzburg-Landau theory still gives reasonable results while the share of the
superconducting electrons is already of order unity so that the screening by
normal particles may be neglected. Previously, it was predicted that the
vortex core in bulk type-II samples is negatively charged. Here we found
that even in the Meissner state with no vortices inside the sample there
exists a non-uniform charge distribution. Due to the finite radial size a
region near the sample edge becomes negatively charged while the interior of
the sample has a corresponding positive charge. This charge redistribution
is a consequence of the screening currents near the sample edge which makes
it behave like a vortex which is turned inside out. When vortices are inside
the sample there is a superposition of the vortex charge and this Meissner
charge. Because of this interplay between vortex charge, which is positive
near the sample surface, and the Meissner charge, which is negative at the
sample surface, the charge at the sample edge changes sign as a function of
the applied magnetic field. We also proved analytically that the there is
only a redistribution of charge and that the total sample charge is neutral
as long as the boundary condition~(\ref{BCdim}) is satisfied.

\section{ACKNOWLEDGMENTS}

This work is supported by the Flemish Science Foundation (FWO-Vl), the
Belgian Inter-University Attraction Poles (IUAP-VI), the ``Onderzoeksraad
van de Universiteit Antwerpen'', and the ESF programme on Vortex Matter.

\section{APPENDIX: PROOF OF ELECTRICAL NEUTRALITY\ IN A MESOSCOPIC\ SAMPLE}

Using Eqs.~(\ref{pot1}-\ref{Poiss}) and Gauss theorem the total charge $Q$
can be expressed as 
\begin{eqnarray}
Q &=&-\varepsilon \frac{\left| \alpha \right| }{2e}\int_{V}\nabla ^{2}\left|
\psi \left( \overrightarrow{r}\right) \right| ^{2}dV  \eqnum{A.1}
\label{App1} \\
&=&-\varepsilon \frac{\left| \alpha \right| }{2e}\int_{S}\overrightarrow{%
\nabla }\left| \psi \left( \overrightarrow{r}\right) \right| ^{2}d%
\overrightarrow{S}  \nonumber \\
&=&-\varepsilon \frac{\left| \alpha \right| }{2e}\int_{S}\overrightarrow{n}%
\cdot \left[ \psi ^{\ast }\left( \overrightarrow{r}\right) \overrightarrow{%
\nabla }\psi \left( \overrightarrow{r}\right) +\psi \left( \overrightarrow{r}%
\right) \overrightarrow{\nabla }\psi ^{\ast }\left( \overrightarrow{r}%
\right) \right] dS  \nonumber \\
&=&-\varepsilon \frac{\left| \alpha \right| }{2e}\int_{S}\overrightarrow{n}%
\cdot \left[ \psi ^{\ast }\left( \overrightarrow{r}\right) \left( 
\overrightarrow{\nabla }-i\overrightarrow{A}\right) \psi \left( 
\overrightarrow{r}\right) +\psi \left( \overrightarrow{r}\right) \left( 
\overrightarrow{\nabla }+i\overrightarrow{A}\right) \psi ^{\ast }\left( 
\overrightarrow{r}\right) \right] dS.  \nonumber
\end{eqnarray}
From the boundary condition~(\ref{BCdim}) and its complex conjugate it
follows immediately that $Q\equiv 0$. Notice, that this result is very
general, it is valid for any vortex configuration (the giant vortex states
and the multi-vortex ones) and arbitrary shape of the superconducting sample
as long as the boundary condition~(\ref{BCdim}) is satisfied.

\begin{figure}[tbp]
\caption{The radial dependences of the Cooper pair density $\left| \protect%
\psi \left( \protect\rho \right) \right| ^{2}$, the charge density $q\left( 
\protect\rho \right) $ and the supercurrent density $j\left( \protect\rho %
\right) $\ in the Meissner state for (a)~an infinite long cylinder with $%
\protect\kappa =1.0$, (b)~finite thickness disk with $d/\protect\xi =0.2$, $%
\protect\kappa =1.0$, (c)~very thin disk for different magnetic fields, and
(d)~for an infinite long cylinder with different Ginzburg-Landau parameter $%
\protect\kappa $ at the magnetic field $H=0.52H_{c2}$. All samples have the
same radius $R/\protect\xi =4.0$.}
\label{L0}
\end{figure}
\begin{figure}[tbp]
\caption{The position of the boundary between the regions of negative and
positive charges inside the sample as a function of the external magnetic
field for: (a)~the infinite long cylinder with $\protect\kappa =1.0$,
(b)~the finite thickness disk with $d/\protect\xi =0.2$, $\protect\kappa =1.0
$, and (c)~the thin disk. The $\protect\rho ^{\ast }\left( H\right) $
dependences for the Meissner state are shown by the solid line, and for the
vortex state - by the dashed lines. The dash-dotted curves represent the
position of zero current.}
\label{rstar}
\end{figure}
\begin{figure}[tbp]
\caption{The absolute value of the negative charge inside the sample as a
function of the external magnetic field for: (a)~the infinite long cylinder
with $\protect\kappa =1.0$, (b)~the finite thickness disk with $d/\protect\xi
=0.2$, $\protect\kappa =1.0$, and (c)~the thin disk. The dependences for the
Meissner state are shown by the solid line, and for the vortex state - by
the dashed lines.}
\label{quabs}
\end{figure}
\begin{figure}[tbp]
\caption{The same as in Fig.~\ref{L0} but for the single vortex state (with $%
L=1$).}
\label{L1}
\end{figure}
\begin{figure}[tbp]
\caption{The radial dependences of the Cooper pair density $\left| \protect%
\psi \left( \protect\rho \right) \right| ^{2}$, the charge density $q\left( 
\protect\rho \right) $ and the supercurrent density $j\left( \protect\rho %
\right) $\ in the giant vortex state with (a)~$L=2$ and (b)~$L=3$ for a thin
disk with $R/\protect\xi =4.0$ and for different applied magnetic fields.}
\label{L23}
\end{figure}
\begin{figure}[tbp]
\caption{The charge density distribution for the ($1$:$2$) saddle-point
state at the field $H=0.32H_{c2}$ for the thin disk with $R/\protect\xi =4.0$%
. The bottom contour plot shows the distribution of the Cooper pair density.}
\label{mv_saddle}
\end{figure}
\begin{figure}[tbp]
\caption{The same as in Fig.~\ref{mv_saddle} but for the ($0$:$4$)
multivortex state in the disk with $R/\protect\xi =4.0$\ at the field $%
H=0.75H_{c2}$.}
\label{mv_stable}
\end{figure}


\begin{references}
\bibitem[*]{perm}  Permanent address: Donetsk Physical \& Technical
Institute, National Academy of Sciences of Ukraine, Donetsk 83114, Ukraine.

\bibitem[{{*}}*]{aut2}  Electronic address: francois.peeters@ua.ac.be

\bibitem{charge1}  D.~I.~Khomskii and A.~Freimuth, Phys. Rev. Lett. {\bf 75}%
, 1384 (1995).

\bibitem{charge2}  G.~Blatter, M.~Feigel'man, V.~Geshkenbein, A.~Larkin, and
A.~van~Otterlo, Phys. Rev. Lett. {\bf 77}, 566 (1996).

\bibitem{charge3}  J.~Kol\'{a}\v{c}ek, P.~Lipavsk\'{y}, and E.~H.~Brandt,
cond-mat/0008299.

\bibitem{Mosch1}  V.~V.~Moshchalkov, L.~Gielen, C.~Strunk, R.~Jonckheere,
X.~Qiu, C.~van~Haesendonck, and Y.~Bruynseraede, Nature (London) {\bf 373},
319 (1995).

\bibitem{Schw1}  V.~A.~Schweigert and F.~M.~Peeters, Phys.\ Rev. B {\bf 60, }%
3084 (1999).

\bibitem{Geim1}  A.~K.~Geim, I.~V.~Grigorieva, S.~V.~Dubonos, J.~G.~S.~Lok,
J.~C.~Maan, A.~E.~Filippov, and F.~M.~Peeters, Nature (London) {\bf 390},
259 (1997).

\bibitem{Deo1}  P.~S.~Deo, V.~A.~Schweigert, F.~M.~Peeters, and A.~K.~Geim,
Phys. Rev.\ Lett. {\bf 79}, 4653 (1997).

\bibitem{Schw2}  V.~A.~Schweigert and F.~M.~Peeters, Phys. Rev. B {\bf 57},
13817 (1998).

\bibitem{Schw3}  V.~A.~Schweigert, F.~M.~Peeters, and P.~S.~Deo, Phys. Rev.
Lett. {\bf 81}, 2783 (1998).

\bibitem{Schw4}  V.~A.~Schweigert and F.~M.~Peeters, Phys. Rev. Lett. {\bf 83%
}, 2409{\bf \ }(1999).

\bibitem{Pal}  J.~J.~Palacios, Physica B {\bf 256-258}, 610 (1998); Phys.
Rev. B {\bf 58}, R5948 (1998); Phys. Rev. Lett. {\bf 84,} 1796 (2000).

\bibitem{Akkerm}  E.~Akkermans and K.~Mallick, J. Phys. A {\bf 32}, 7133
(1999); E.~Akkermans, D.~M.~Gangardt, and K.~Mallick, Phys. Rev. B {\bf 62},
12427 (2000); cond-mat/0008289.

\bibitem{Deo2}  P.~S.~Deo, F.~M.~Peeters, and V.~A.~Schweigert,
Superlattices Microstruct. {\bf 25}, 1195 (1999).

\bibitem{cyl1}  H.~J.~Fink and A.~G.~Presson, Phys. Rev. {\bf 151}, 219
(1966); Phys. Rev. {\bf 168}, 399 (1968).

\bibitem{Zharkov}  G.~F.~Zharkov, V.~G.~Zharkov, and A.~Yu.~Zvetkov, Phys.
Rev. B {\bf 61}, 12293 (2000); cond-mat/0008217; G.~F.~Zharkov,
cond-mat/0010028.

\bibitem{ring1}  H.~J.~Fink and V.~Gr\"{u}nfeld, Phys. Rev.\ B {\bf 33},
6088 (1986); A.~Bezryadin, A.~Buzdin, and B.~Pannetier, Phys. Rev. B {\bf 51}%
, 3718 (1995); E.~M.~Horane, J.~I.~Castro, G.~C.~Buscaglia, and
A.~L\'{o}pez, Phys. Rev. B {\bf 53}, 9296 (1996); V.~Bruyndoncx, L.~Van
Look, M.~Verschuere, and V.~V.~Moshchalkov, Phys. Rev. B {\bf 60}, 10468
(1999).

\bibitem{Ben1}  B.~J.~Baelus, F.~M.~Peeters, and V.~A.~Schweigert, Phys.
Rev. B {\bf 61}, 9734 (2000); F.~M.~Peeters, V.~A.~Schweigert, B.~J.~Baelus,
and P.~S.~Deo, Physica C {\bf 332}, 255 (2000).

\bibitem{FK}  V.~M.~Fomin, V.~R.~Misko, J.~T.~Devreese, and
V.~V.~Moshchalkov, Phys. Rev. B {\bf 58}, 11703 (1998).

\bibitem{book1}  P.~G.~de~Gennes, {\it Superconductivity of Metals and
Alloys }(Addison-Wesley, New York, 1994).

\bibitem{Schw5}  V.~A.~Schweigert and F.~M.~Peeters, Physica C {\bf 332},
266 (2000).

\bibitem{tamm}  I.~E.~Tamm, {\it Fundamentals of the Theory of Electricity}
(Mir Publishers, Moscow, 1979), p.~66.

\bibitem{enhance}  S.~V.~Yampolskii and F.~M.~Peeters, Phys. Rev. B {\bf 62}%
, 9663 (2000).

\bibitem{book2}  {\it Handbook of Mathematical Functions}, eds. by
M.~Abramovitz and I.~Stegun (Dover Publications, New York, 1970), p.~504.

\bibitem{Ben2}  B.~J.~Baelus, F.~M.~Peeters, and V.~A.~Schweigert,
cond-mat/0010217.
\end{references}
\end{document}